\documentclass[aps,prd,twocolumn,showpacs,eqsecnum,nofootinbib,floatfix]{revtex4-1}
\pdfoutput=1

\usepackage{amssymb}
\usepackage{gensymb}
\usepackage{graphicx}
\usepackage{amsmath}
\usepackage{wasysym}
\usepackage{verbatim}

\font\FermiSmallfont=cmssq8 scaled 1200

\def\UMDppthead#1#2#3{
\null 
\begin{center}\vskip -1.0truein{\hbox to 7.5truein {
\hfill
\vbox to 1in {\vfill \FermiSmallfont
              \hbox{#1}
              \hbox{#2}
              \hbox{#3}
              \vfill}
}}\vskip-0.0truein\end{center}}%FNALppthead

\def\apjs{Astrophys.\ J.\ Supp.}
\def\apj{Astrophys.\ J.}

\def\apjl{Astrophys.\ J.\ Lett.}
\def\aap{Astron.\ \& Astrophys.}
\def\mnras{Mon. Not. Roy. Astron. Soc.}

\def\pasp{Pub. Astron. Soc. Pacific}
\def\aaps{Astron.\ \& Astrophys.\ Supp.}

\begin{document}

\UMDppthead{UMD-PP-10-022}{arXiv:1012.1247}{}

\title{The Contribution of Blazars to the Extragalactic Diffuse
  Gamma-ray Background and Their Future Spatial Resolution}

\author{Kevork N.\ Abazajian$^{1,2}$} \email{kevork@uci.edu}
\author{Steve Blanchet$^{2,3}$} \email{steve.blanchet@uam.es}
\author{J.\ Patrick Harding$^2$} \email{hard0923@umd.edu}
\affiliation{$^1$Department of Physics and Astronomy, University of
  California, Irvine, Irvine, California 92697 USA}
\affiliation{$^2$Maryland Center for Fundamental Physics \& Joint
  Space-Science Institute, Department of Physics, University of
  Maryland, College Park, Maryland 20742 USA}
\affiliation{$^3$Instituto de F\'isica Te\'orica, IFT-UAM/CSIC Nicolas
  Cabrera 15, UAM Cantoblanco, 28049 Madrid, Spain} \date{\today}

\begin{abstract}
  We examine the constraints on the luminosity-dependent density
  evolution model for the evolution of blazars given the observed
  spectrum of the diffuse gamma-ray background (DGRB), blazar
  source-count distribution, and the blazar spectral energy
  distribution sequence model, which relates the observed the blazar
  spectrum to its luminosity.  We show that the DGRB observed by the
  Large Area Telescope (LAT) aboard the {\it Fermi Gamma Ray Space
    Telescope} can be produced entirely by gamma-ray emission from
  blazars and nonblazar active galactic nuclei, and that our blazar
  evolution model is consistent with and constrained by the spectrum
  of the DGRB and flux source-count distribution function of blazars
  observed by Fermi-LAT.  Our results are consistent with previous
  work that used EGRET spectral data to forecast the Fermi-LAT DGRB.
  The model includes only three free parameters, and forecasts that
  $\gtrsim 95\%$ of the flux from blazars will be resolved into point
  sources by Fermi-LAT with 5 years of observation, with a
  corresponding reduction of the flux in the DGRB by a factor of
  $\sim$2 to 3 (95\% confidence level), which has implications for the
  Fermi-LAT's sensitivity to dark matter annihilation photons.
\end{abstract}
\maketitle

\section{Introduction}

The source of the extragalactic isotropic diffuse gamma-ray background
(DGRB) has been an unsolved question in astrophysics for some time. In
this paper, we show how the DGRB spectrum can be produced by a
combination of blazar and nonblazar active galactic nuclei (AGN)
gamma-ray sources. We also show that the blazar flux source-count
distribution function ($dN/dF$) is consistent with the full DGRB
originating from these sources. Furthermore, we show how less-detailed
models of the blazar contribution failed to be consistent with the
DGRB.  We explore how the implications for dark matter detection or
constraints from the DGRB will evolve as the blazar sources of the
DGRB are resolved.

The DGRB was first discovered by the SAS 2 experiment in 1975, for
gamma-ray emission in the range of 35 to 300
MeV~\cite{Fichtel75,*Fichtel77,*Fichtel78,*Thompson82}. This background
was seen at energies up to 20 GeV by the EGRET Collaboration, and it
was confirmed at these energies in the first-year data from the Large
Area Telescope (LAT) aboard the {\it Fermi Gamma-Ray Space
  Telescope}~\cite{Sreekumar98,Abdo10b,Abdo10}. The assumed
extragalactic source of the DGRB is determined by measuring the
complete diffuse (unresolved) flux and then subtracting off a model to
account for the background coming from our Galaxy. This yields a
measure of the flux coming from unresolved diffuse sources, presuming
there is no minimal isotropic component from the Galaxy, e.g. dark
matter annihilation or decay. The DGRB has been used to constrain dark
matter annihilation in Galactic and extragalactic
sources~\cite{Abdo10c,Abazajian10,Baltz08}.

The most recent measurement of the DGRB was performed by the
Fermi-LAT. In the Fermi-LAT Collaboration analysis, the gamma-ray
intensity was measured in the range of 100 MeV to 100 GeV above
$10\degree$ in Galactic latitude ($\lvert b\rvert >10\degree$). The
total flux is modeled by stacking the spectra of known sources with
the cosmic-ray background, the Galactic diffuse background, and the
DGRB. This analysis gives a DGRB intensity that is roughly 25\%
of the total observed flux. The DGRB seen by the Fermi-LAT is
consistent with a power law in energy with index 2.41. This value for
the DGRB is notably softer at high energies than was previously seen
in the EGRET Collaboration, which is partly due to an updated model of
the diffuse Galactic emission in Ref.~\cite{Abdo10b} (hereafter FS10).

A detailed spectral energy distribution (SED) sequence model of
blazars can reproduce the DGRB~\cite{Inoue09,Inoue10a}. We explore
this model in this work. Many models have been proposed to explain the
DGRB. It has been shown that emission from AGN can account for the
diffuse background from 10 keV to 100 MeV, but above that energy, this
model cannot account for the large gamma-ray
flux~\cite{Inoue08}. Radiation from star-forming galaxies could
account for much of the DGRB up to 10 GeV, but this also cannot
explain the high intensities observed at higher
energies~\cite{Fields10}. Emission from millisecond pulsars has been
proposed as a source as well~\cite{FaucherGiguere09}. However,
millisecond pulsars as a dominant source of the DGRB may be
inconsistent with the lack of anisotropy in the
DGRB~\cite{SiegalGaskins10}.

Dark matter annihilation, both as a component of the extragalactic
diffuse emission and as an unaccounted foreground from the Milky Way
can contribute to the DGRB, but the fluxes from dark matter are
expected to be lower than the DGRB flux and have a different spectral
shape~\cite{Abazajian10,Abdo10c}. However, measurements of the DGRB
are one of the strongest ways to constrain dark matter
annihilation~\cite{Baltz08}.  If dark matter is a significant
contributor, it may be disentangled from astrophysical sources due to
its angular correlation on the
sky~\cite{Ando:2005xg,*Ando:2006cr,*Miniati07,*SiegalGaskins:2008ge,*SiegalGaskins:2009ux,*Hensley09,*Fornasa:2009qh}. Pioneering
work proposed that blazars could account for {\it all} of the DGRB
seen by the EGRET Collaboration~\cite{Stecker96}.  The blazar class of
AGN has been studied in depth as the origin of the DGRB at high
energies~\cite{Stecker93,Stecker96,Padovani93,Salamon94,Chiang95,Chiang98,Mucke00,Giommi05,Narumoto06,Dermer07,Pavlidou08,Bhattacharya09}.

In Ref.~\cite{Inoue09} it was shown that the DGRB can be composed of
blazars and nonblazar AGN in the luminosity-dependent density
evolution (LDDE) SED blazar model. This
model contains only three free parameters describing the gamma-ray
luminosity function (GLF) of blazars. We show that this model is
consistent with producing the full DGRB spectrum as well as the blazar
source-count distribution, $dN/dF$, of blazars as measured by
Fermi-LAT.  In addition, we constrain this model by these measurements
and find parameters for which the model successfully reproduces these
measurements. Note that both the source-count distribution $dN/dF$ and
DGRB spectrum are predicted by the model, and not an input to the model.

Recent work by the Fermi-LAT Collaboration found that the DGRB could
not be composed entirely by blazars~\cite{Abdo10} (hereafter FB10).
However, that work adopted an over-simplification of the blazar SED to
be a single power-law (PL), independent of blazar luminosity, which is
inconsistent with the observed spectral luminosity dependence seen in
the SED sequence~\cite{Fossati97,*Fossati98,*Donato01}.  In contrast,
in a separate paper, the Fermi-LAT Collaboration emphasizes the need
for including departures from pure-PL behavior in blazar spectra when
calculating the contribution of unresolved low-luminosity blazars to
the DGRB~\cite{Abdo10a}.  Incorporating the SED departure and its
dependence on blazar luminosity evolution when modeling the DGRB is
exactly the intent of the work presented here.

Furthermore, the blazar model in FB10 lacks a physical evolution model
for blazars.  Instead of the source-count distribution resulting from
the cosmological evolution of blazars, the source-count distribution
is an input to the model, as a broken power-law with four free
parameters.  Note that even though the model in FB10 is simplistic, it
contains {\it more} free parameters than the LDDE plus SED-sequence
model explored here.  In our approach there are three parameters in
the adopted blazar model which describe the relation between the GLF
and x-ray luminosity function (XLF).  Because the FB10 model employs a
pure-PL luminosity-independent SED with a broken-PL source-count
distribution, the conclusions of that work do not apply to the model
examined here.  Other parameters in our work ({\it e.g.}, the SED
sequence and the low-energy nonblazar AGN model) are constrained by
other observations and remain fixed in our blazar model analysis.
Namely, the observational constraints on the SED sequence come from
spectral population models of blazars as in
\cite{Fossati97,*Fossati98,*Donato01}, and the nonblazar AGN spectrum
is constrained by the hard x-ray luminosity function derived from {\em
  HEAO1}, {\em ASCA}, and {\em Chandra} x-ray AGN surveys
\cite{Ueda03,Inoue08}.

A recent paper by Malyshev and Hogg~\cite{Malyshev:2011zi} using the
one-point probability distribution function (PDF) of the DGRB also
concludes that blazars cannot constitute the total DGRB flux as
measured by Fermi-LAT, when modeled as a pure-PL SED with a fixed
$dN/dF$.  However, this conclusion also only applies to the model
which they consider, which adopt blazars as having pure-PL
luminosity-independent SEDs, and not to the LDDE SED-sequence model
examined here.

Because observed blazars make up about 15\% of the total
gamma-ray flux, unresolved blazars are a likely candidate to make up
the DGRB~\cite{Sreekumar98,Abdo10}. Blazars were the most numerous
point-source objects observed by the EGRET
Collaboration~\cite{Hartman99}. Additionally, observed blazar spectra
tend to follow a similar power law in energy as the DGRB.  However, it
is known that blazars have a luminosity dependence to their spectral
shape, which is incorporated in the SED-sequence model
\cite{Fossati97,*Fossati98,*Donato01}, but ignored in the analysis of FB10.

Blazars are the combination of two classes of AGN: flat-spectrum
radio quasars (FSRQs) and BL Lacertae objects (BL Lacs). FSRQs are
AGN that have spectral index $\alpha_{r}<0.5$ in the radio band and
have radio emission lines with equivalent width greater than $5\rm\
\AA$. BL Lacs have no strong absorption or emission features, and have
equivalent widths less than $5\rm\ \AA$~\cite{Urry95}. Broadly
speaking, blazars tend to have their bolometric luminosities dominated
by the gamma-ray luminosity and have great variability in that
luminosity. Therefore, it is believed that blazars represent the small set
of AGN that are observed along the jet axis, as opposed to nonblazar
AGN which are observed far from the jet axis and dominate emission by
their luminous accretion disk. This jet source is expected to be
relativistically beamed, as opposed to the more isotropic flux coming
from the AGN's accretion disk~\cite{Blandford79,Dermer95}.

Different models of blazar emission have been proposed in the
literature~\cite{Padovani93,Stecker93,Salamon94,Stecker96,Mucke00,Giommi05,Pavlidou08,Chiang95,Chiang98,Dermer07,Bhattacharya09,Narumoto06}. One
is the pure luminosity evolution (PLE) model of the distribution of
blazars~\cite{Chiang95,Chiang98,Dermer07,Bhattacharya09}. In this
model, only the blazar luminosity is evolved in redshift. An
alternative model, LDDE, relates the gamma-ray luminosity of blazars
to the redshift-dependent distribution of x-ray emission from
nonblazar AGN~\cite{Narumoto06}. This technique more realistically
fits the blazar evolution to the AGN distribution, rather than
assuming that all blazars have identical evolution regardless of
luminosity. In many models for blazar spectra, a simple power-law or
distribution of power laws is used as the intrinsic blazar spectrum,
but more detailed frequency-dependent models have been used as
well~\cite{Giommi05}.

Here, we employ the LDDE model for blazar distributions. For the
intrinsic spectrum of blazars, we use a frequency-dependent SED based
on the multiwavelength study of
Ref.~\cite{Fossati97,*Fossati98,*Donato01}. We use these models to
derive the differential blazar spectrum in redshift, luminosity, and
energy. By integrating over these variables, we can determine the
number of detectable blazars for given detector sensitivities, and we
can calculate the expected gamma-ray flux from unobserved blazars to
determine how significantly they contribute to the DGRB. Additionally,
we add a fixed nonblazar AGN component to our predicted blazar flux,
which should make the net flux from our model fit the diffuse
background over the energy range from 10 keV to 100 GeV.

Below, we begin by describing the DGRB seen by the Fermi-LAT as well
as its data on blazars. We will then describe our model in detail,
specifying the evolution model and SED used in our calculations and
how we fit these to the known data. We use this model to predict the
ability of the Fermi-LAT to detect blazars and how this will affect
the DGRB. Throughout the paper, we take a flat universe with the
cosmological parameters $\Omega_{m}=0.272$, $\Omega_{\Lambda}=0.728$,
and $H_0=70.2\rm\ km\ s^{-1}\ Mpc^{-1}$~\cite{Komatsu10}. {\it Note,}
the use of $h$ in the text refers to Planck's constant, and not the
Hubble parameter.

\section{First-Year Findings by the Fermi-LAT Collaboration}

\subsection{DGRB Measurements}

From its first year of data, the Fermi-LAT has measured a spectrum for
the DGRB (FS10). To get this spectrum, the total gamma-ray
intensity had known sources subtracted from it, as well as the
background from cosmic rays, and the expected Galactic diffuse
emission. At this time, resolved extragalactic sources account for
about 15\% of the total gamma-ray flux in the sky.  To calculate
the gamma-ray emission from Galactic cosmic rays, the local cosmic-ray
spectra are extrapolated to give source populations, which are then
propagated through appropriate target distributions using the GALPROP
particle propagation package \cite{Strong98,Strong00}. This diffuse
Galactic emission is the largest component of the DGRB, comprising
roughly half of the total observed intensity. A small component to the
DGRB is a background due to cosmic-ray interaction with the Fermi-LAT
itself. This background has been studied in detail in
FS10 and is very well characterized. This background
accounts for 1 to 10\% of the total emission, with a greater
fraction at low energies and a lesser fraction at high energies. The
residual intensity after all of these components have been removed is
called the isotropic DGRB.  It makes up around 25\% of the total
emission. Because of the model dependence of these subtractions, the
uncertainties on the DGRB are dominated by systematics (FS10).
The DGRB may come from unresolved extragalactic sources or unaccounted
Galactic sources, such as millisecond pulsars, or, potentially, from
Galactic dark matter annihilation or decay. 

\subsection{Point-Source Sensitivity}

\begin{figure}[t]
\begin{center}
\includegraphics[width=3.4truein]{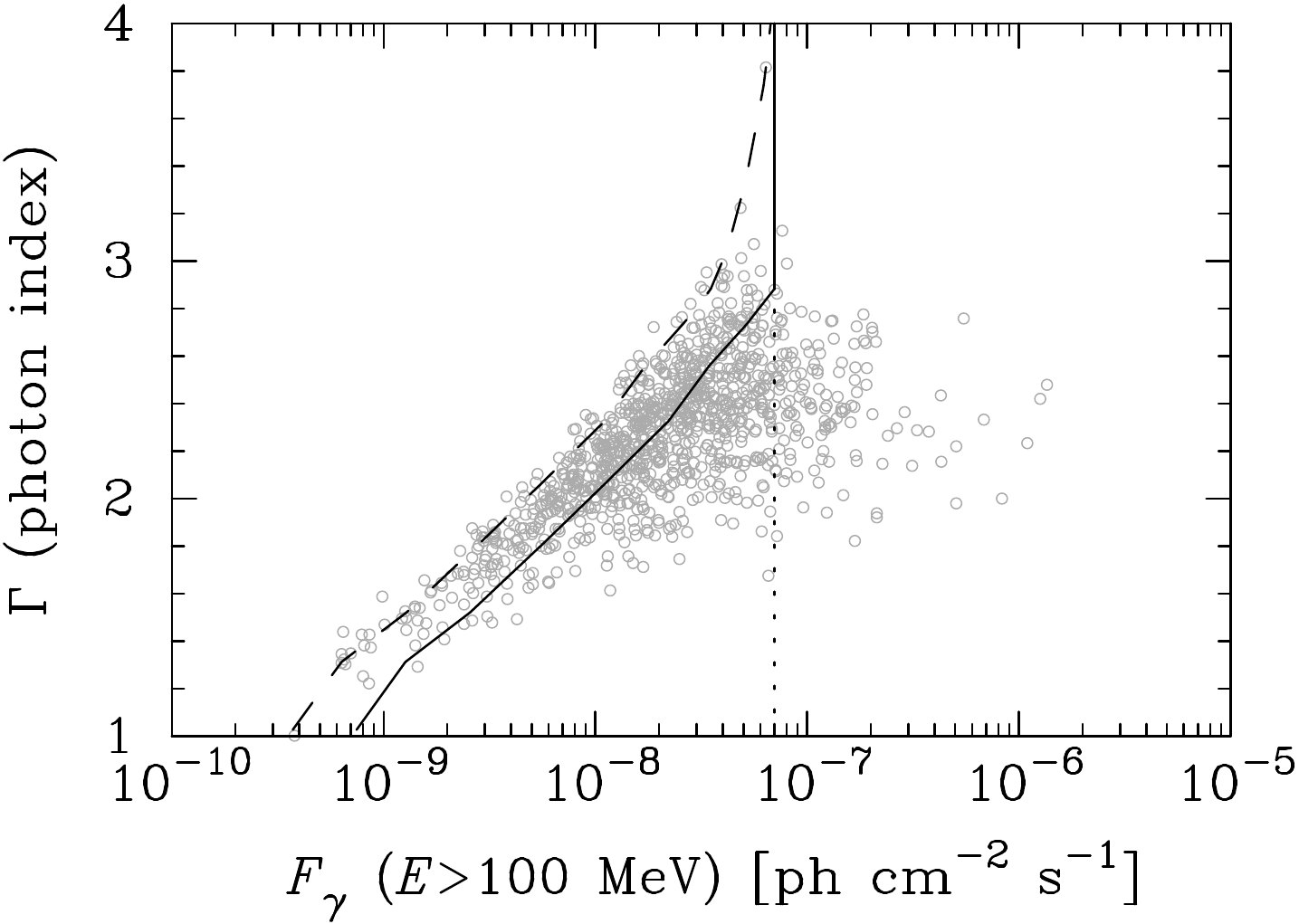}
\end{center}
\caption{Shown is a sample of the blazar gamma-ray fluxes above 100
  MeV ($F_{100}$) versus their power-law fit spectral-index $\Gamma$ from FB10.
  The blazars (points) are shown above point-source detection
  test-statistic $TS=25$ (with the corresponding point-source limit
  shown as the dashed line), while those below $TS=50$ are modeled, in
  our work and in FB10, to contribute to the DGRB as measured by FS10
  (with point-source limit shown by the solid line). Note the
  total luminosity vs spectrum dependence of the blazar population evident
  in this plot.  The first-year Fermi-LAT point-source sensitivity is
  complete above the dashed line at $7\times 10^{-8}\rm\ ph\
  cm^{-2}s^{-1}$ (FB10). \label{point}}
\end{figure}

The Fermi-LAT detector has a spectrally dependent point-source
sensitivity due to the higher spatial resolution of the instrument to
higher-energy photons.  The flux limit to point sources is shown in
Fig.~\ref{point}, along with the sample of blazar fluxes and spectral
indices from FB10.  In FS10, the DGRB spectrum is compared to that
measured by EGRET, which had a point-source sensitivity of $1\times
10^{-7}\rm\ ph\ cm^{-2}s^{-1}$, despite the fact that the point-source
sensitivity of the two instruments, and therefore the measured DGRB
flux between the two instruments' measurements, are quantitatively
different.\footnote{Because of this direct comparison in FS10, in the v1
  preprint of this work, a point-source sensitivity cutoff of the
  measured DGRB spectrum of FS10 was adopted to be $1\times
  10^{-7}\rm\ ph\ cm^{-2}s^{-1}$, instead of the spectrally dependent
  sensitivity here.  This does not change our conclusions, but does
  modify our best-fit model parameters and our 5-year forecast DGRB
  spectra.}  We derive the flux limit from the sample of blazars used
in FB10, using the lowest-flux end of the blazar sample, which
satisfied the test-statistic $TS=25$.  In FB10, the source-count
distribution and DGRB spectrum was fit with only blazars resolved at
$TS=50$; therefore, the point-source limit is augmented by a factor of
2, as shown by the solid in Fig.~\ref{point}, with the point-source
sensitivity always below or equal to Fermi-LAT's believed completeness
for all spectra sources at $7\times 10^{-8}\rm\ ph\ cm^{-2}s^{-1}$.

Importantly, it should be made clear that a fixed point-source
sensitivity cannot be exactly specified for the DGRB spectrum derived
in FS10.  In that work, all sources above a $TS=200$ are allowed to
vary in the amplitude of their flux during the fitting of the
extragalactic isotropic DGRB.  Therefore, the exact flux-limit of the
DGRB spectrum, and therefore the nature of the spectrum itself, as
presented in FS10, is ill-defined.  We therefore adopt the
best-estimate method of modeling the DGRB spectrum as done by the
Fermi-LAT Collaboration itself in FB10, with a $TS=50$
spectrally-dependent flux limit.  We define the power-law photon index
$\Gamma$ for the non-power-law SED-sequence model of a blazar by
fitting a power law to the Poisson-limited spectrum within the
observed energy range of Fermi-LAT.

As the point-source sensitivity of Fermi-LAT improves with integration
time, the resolution of the extragalactic DGRB into point sources will
not proceed proportionally to the sensitivity, but rather in a
combination of the sensitivity with where the population of
extragalactic emitters lies with respect to that
sensitivity/spectral-index plane.  In particular, for the LDDE plus
SED-sequence blazar model here, there are more hard-spectrum sources with
lower gamma-ray flux.  This trend already can be seen in the plotted blazar
points in Fig.~\ref{point}.

\subsection{Blazar Measurements}

Through one year of running, the Fermi-LAT has detected a total of 296
FSRQs, 300 BL Lacs, and 72 blazars of unknown type. The observed FSRQs
have an average spectrum with photon index 2.48 and BL Lacs have
average photon index of 2.07~\cite{Abdo10d}. This power-law index is
similar to the DGRB power-law index of 2.41, which suggests that
unresolved blazars could be the primary source of the
DGRB. Additionally, the stacked spectra of known blazars detected by
the Fermi-LAT are responsible for 15\% of their total observed
gamma-ray emission observed by the Fermi-LAT. The number of blazars
observed above a given flux tends to follow a broken power law, with a
break at $F(>100\rm\ MeV)=6\times 10^{-8}\rm\ photons\ cm^{-2}\
s^{-1}$. This break seems to be independent of detector sensitivity,
because the sensitivity dies off much more quickly as a function of
flux than the blazar number count (FB10).

In the Fermi-LAT measurements, FSRQs and BL Lacs have similar
variability properties, so the assumption that they are of one class
appears valid. For BL Lacs, the LAT has detected significantly more
hard-spectrum sources than soft-spectrum sources, which is consistent
with the known selection bias in the measurement. FSRQs peak at a
redshift of unity, indicating that the sample is approaching
completeness. In contrast, BL Lacs peak at low redshift, indicating
that the sample is not yet complete. FSRQs tend to be more luminous
than BL Lacs: FSRQs have radio luminosities that peak at $L_{\rm
  rad}\approx 10^{44.5}\rm\ erg/s$ whereas BL Lacs have lower radio
luminosities peaking at $L_{\rm rad}\approx 10^{42}\rm\
erg/s$~\cite{Abdo10d}. This would indicate that there is a fairly
large contribution of low-luminosity, soft-spectrum BL Lacs that has
yet to be resolved.

The differences in spectra between FSRQs and BL Lacs are
significant. The average gamma-ray photon index is roughly 0.5 larger
for FSRQs than for BL Lacs. Even among BL Lacs themselves,
high-synchrotron-peak BL Lacs have a photon index of 2.28 while
low-synchrotron-peak BL Lacs have a photon index of 1.96. FSRQs give
off their peak synchrotron radiation at around $10^{13}\rm\ Hz$
whereas for BL Lacs, the distribution is much broader, stretching from
$10^{12}\rm\ Hz$ to $10^{17}\rm\ Hz$~\cite{Abdo10d}. FSRQs have their
inverse Compton (IC) peaks at energies less than 100 MeV, so power-law
fits work fairly well to match their LAT-measured spectra. For BL
Lacs, the peak IC emission tends to lie in the LAT's energy range,
with low-synchrotron-peak BL Lacs peaking closer to 100 MeV and
high-synchrotron-peak BL Lacs peaking closer to 100 GeV. Because of
these peaks, these spectra do not match a power-law, though a broken
power-law can approximately fit them~\cite{Abdo10a}.

To truly model the blazar SED, a multiwavelength analysis is
needed~\cite{Fossati97,*Fossati98,*Donato01}. The Fermi-LAT
Collaboration did a multiwavelength study of the spectra of blazars,
combining the results of several radio, x-ray, optical, and gamma-ray
blazar studies~\cite{Abdo09a}. This study found strong correlation
between the x-ray and gamma-ray spectral slopes, indicating that
blazar spectra fit a two-peaked, synchrotron plus IC scenario
well. They found that BL Lacs have larger synchrotron peaks than
FSRQs, which explains why BL Lacs have harder gamma-ray indices. This
study plotted the SED for several blazars, all of which have a strong
double-peaked shape when luminosity is plotted versus frequency on a
log-log plot. This is consistent with previous analyses of the blazar
SED~\cite{Fossati97,*Fossati98,*Donato01}.

\section{Determination of Blazar Flux and Spectrum}

\subsection{Spectral Energy Distribution}

The model of blazar emission we use consists of two parts: a GLF to
give the density of blazars per unit luminosity and an SED to
determine the luminosity of blazars as a function of energy. These are
denoted by $\rho_{\gamma}(L_{\gamma},z)$ and $\nu L_{\nu}(x;P)$, where
$z$ is redshift of the blazar, $L_{\gamma}$ is the gamma-ray
luminosity (defined as $\nu L_{\nu}$ at $h\nu=100\rm\ MeV$), $x\equiv
\log_{10}(\nu/\rm Hz)$ for blazar rest-frame frequency $\nu$, and $P$
is the bolometric luminosity. Because our SED separates blazars
according to radio luminosity, the bolometric luminosity is used to
determine which SED curve matches a given blazar. For a given SED
curve, the bolometric luminosity can be calculated as $\int
L_{\nu}d\nu$. This can then be used to find the gamma-ray luminosity.

Ref.~\cite{Fossati97,*Fossati98,*Donato01} analyzed the relationship
between frequency and luminosity for blazars. To get these
relationships, blazars were binned by radio luminosity. This analysis
showed that blazar gamma-ray index is correlated with blazar
luminosity. This correlation is consistent with the experimental
results that FSRQs have high luminosities and large gamma-ray spectral
indices while BL Lacs have lower luminosities and smaller spectral
indices~\cite{Ghisellini09,Abdo09a,Abdo10a}. A proper calculation
using blazar spectra should account for this relationship between
index and luminosity, and not simply use a power law in energy for the
blazar spectrum. Note that this was not done in Ref.~\cite{Abdo10},
which claimed that blazars cannot constitute the full DGRB.

For the frequency dependence of the blazar luminosity, we use the SED
sequence of Inoue and Totani~\cite{Inoue09}. In this model, blazars
SEDs are fit over frequencies from radio to gamma ray, as in
Ref.~\cite{Fossati97,*Fossati98,*Donato01}. Each SED is comprised of
two components, a synchrotron component at lower energies and an IC
component at higher energies. These are each parameterized by a
parabolic peak with a lower-energy linear tail. The details of the
model are determined by fitting to the data in
Ref.~\cite{Fossati97,*Fossati98,*Donato01}, which give $\nu L_{\nu}$ as a
function of rest-frame frequency $\nu$ for five luminosity bins. This
provides the gamma-ray luminosity ($\nu L_{\nu}$ at $h\nu=100\rm\
MeV$), the specific luminosity $L_{\nu}(\nu)$, and the bolometric
luminosity $\int L_{\nu}d\nu$ for a blazar with known radio band
luminosity ($\nu L_{\nu}$ at 5 GHz). The full model can be found in
Appendix~\ref{SED Appendix}.

As a check on the versatility of the SED model, we explicitly compared
the model to several blazar spectra measured by the Fermi-LAT
Collaboration~\cite{Abdo10a,Abdo09a}. The model fit the data in the
Fermi-LAT energy range well. It also matched the data qualitatively:
the model spectra had increasing, decreasing, or flat spectral shapes
in agreement with the Fermi-LAT-measured spectra. Such agreement
indicates that this SED fit approximates the full blazar SED well.

\subsection{Gamma-ray Luminosity Function}\label{GLFsection}

For the distribution of gamma-ray blazars, we follow the hard x-ray
AGN distributions parameterized by Ueda et al~\cite{Ueda03}. Similar
work was done for soft x-rays by Hasinger et
al~\cite{Hasinger05}. However, the hard x-ray parameterization gives a
more conservative prediction of blazar detection by Fermi-LAT, so we use
that here. For rest-frame (emission frame) energy of $\epsilon_{\rm
  gam,res}= 100\rm\ MeV$, the gamma-ray luminosity is given by
$L_{\gamma}\equiv (\epsilon_{\rm gam,res}/h)L_{\nu}(\epsilon_{\rm
  gam,res}/h,P)$.

Reference~\cite{Inoue09} argues that the gamma-ray luminosity can be
related to the x-ray AGN disk luminosity $L_{X}$ through the
bolometric luminosity by $P=10^{q}L_{X}$, where $q$ is a scaling
parameter. This is because the bolometric luminosity from a blazar jet
is proportional to the mass accretion rate $\dot{m}$. For blazars
with low accretion rate, the conversion of power into luminosity is
inefficient, with $L_{X}\propto \dot{m}^{2}$. For blazars with high
accretion rate close to the Eddington limit, the conversion is
efficient and the disk luminosity goes as $L_{X}\propto
\dot{m}$~\cite{Falcke03,Merloni03,Gallo05}. Because black hole growth
takes place mostly near the Eddington limit, it is reasonable to
assume that $P\propto\dot{m}\propto L_{X}$~\cite{Marconi03}. Note,
$L_{X}$ is the x-ray luminosity from the accretion disk of the blazar,
not to be confused with the x-ray luminosity of the beam.

The comoving number density per unit $L_{\gamma}$ of gamma-ray blazars
is 
\begin{equation}
\rho_{\gamma}(L_{\gamma},z)=\kappa\frac{dL_{X}}{dL_{\gamma}}\rho_{X}(L_{X},z),
\label{rhogamma}
\end{equation}
where $\rho_{X}$ is the comoving number density of AGN per unit
$L_{X}$, $z$ is the redshift to the source, and $\kappa$ is the
fraction of AGN observed as blazars. The quantity $\rho_{\gamma}$ is
referred to as the GLF. A parameterization of the x-ray luminosity
function $\rho_{X}$ is found in Appendix~\ref{XLF Appendix}. The GLF
has three free parameters: $q$ determines the ratio of bolometric jet
luminosity to accretion-disk x-ray luminosity, $\gamma_{1}$ is the
faint-end index that determines how the GLF behaves for low
luminosities, and the blazar fraction is $\kappa$.

These GLF models are based on LDDE of AGN, as opposed to PLE
models. In PLE models, AGN luminosity changes with redshift, but the
comoving density of AGN remains constant. This has been a popular
method of determining blazar
parameters~\cite{Chiang95,Chiang98,Dermer07,Bhattacharya09}. LDDE
models have a peak evolution redshift which depends on luminosity, so AGN
of different luminosities will have slightly different
evolutions~\cite{Ueda03,Hasinger05}. This gives a better fit to the
AGN data and should describe blazar evolution more fully than PLE
models~\cite{Narumoto06}.  The exact relationship between x-ray AGN
and gamma-rays blazars is not yet known. We are using the simple
ansatz that they are related as shown in Eq.~\eqref{rhogamma}, as
proposed by Inoue and Totani~\cite{Inoue09}. To the best of our
knowledge, this model satisfies all current observations and
constraints, and therefore is a viable possibility.

\begin{figure*}[t]
\begin{center}
\includegraphics[width=2.3truein]{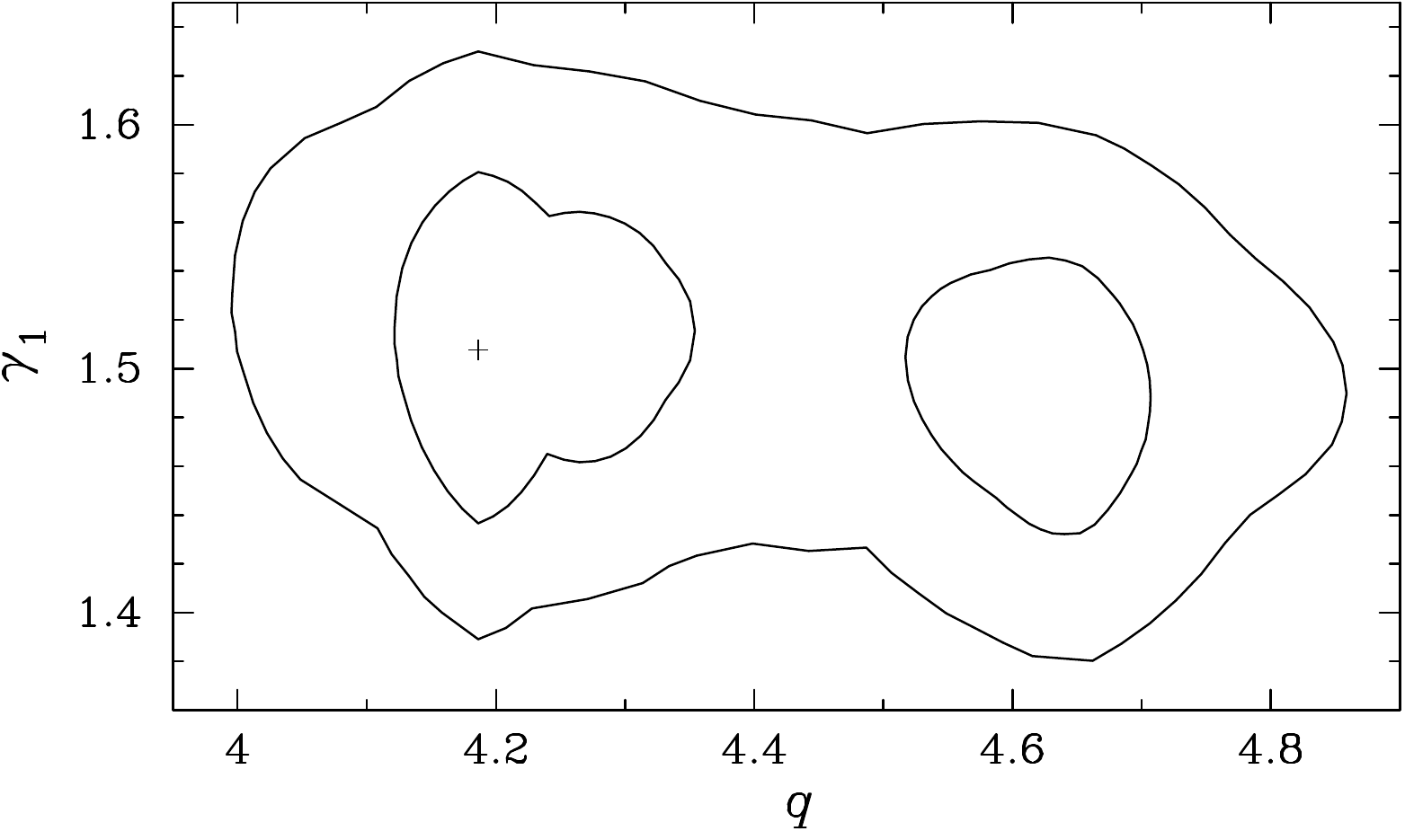}\ \  
\includegraphics[width=2.3truein]{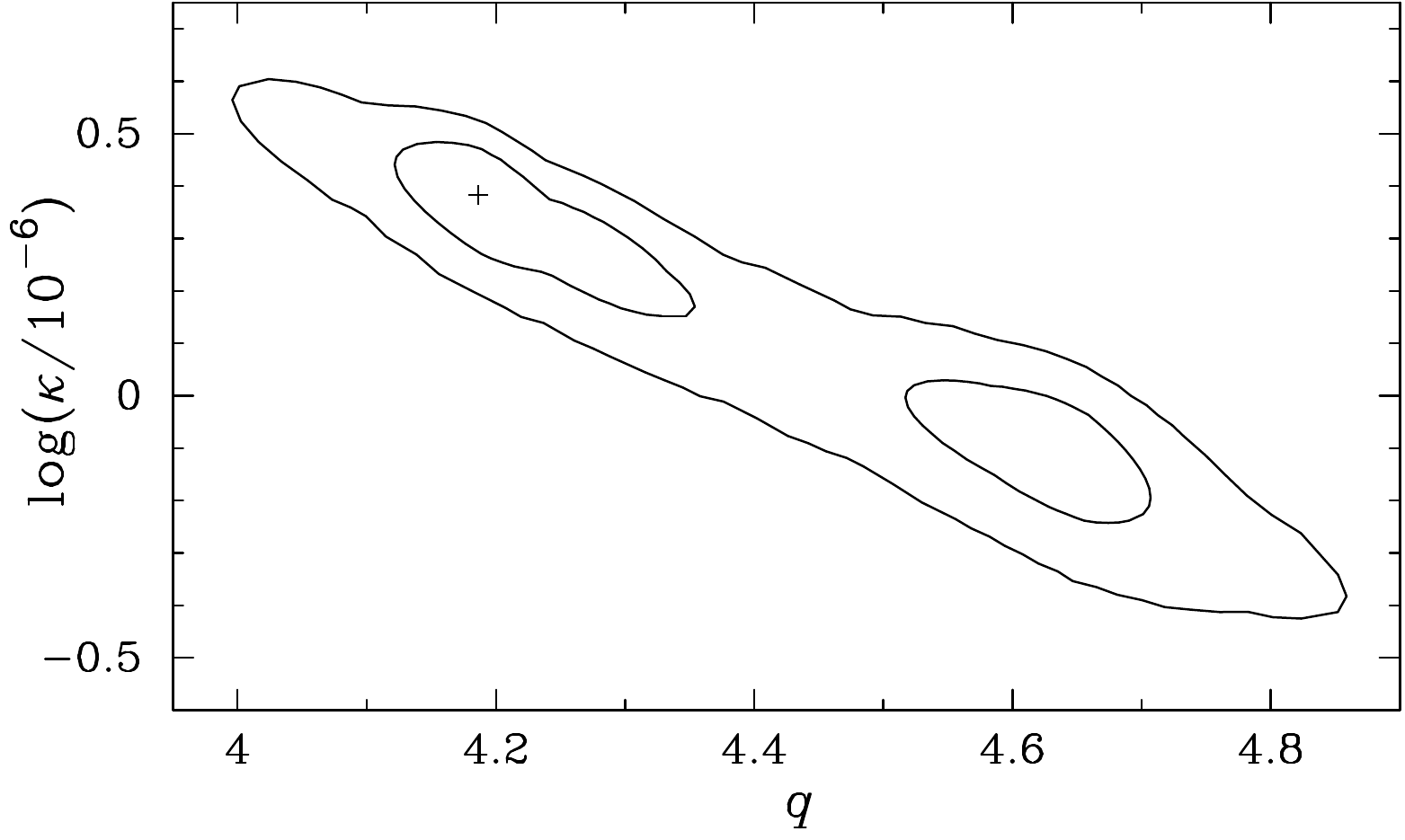}\ \ 
\includegraphics[width=2.3truein]{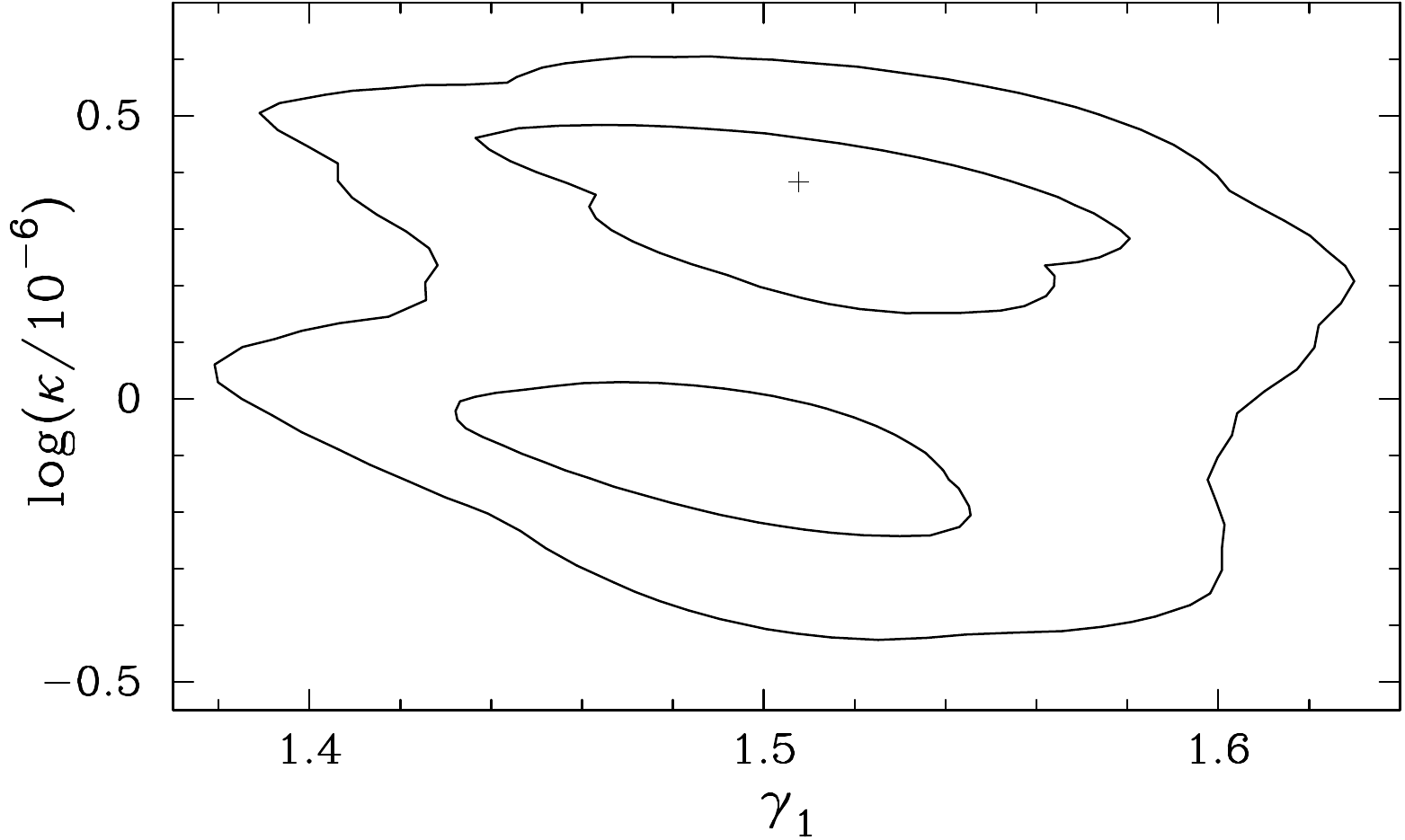}
\end{center}
\caption{Shown are contours with $68\%$ and $95\%$ confidence level
  (CL) regions for the parameters of the luminosity scale $q$ and GLF
  faint-end index $\gamma_{1}$, $q$ vs $\kappa$, and $\kappa$
  vs $\gamma_1$. The best-fit value is labeled by the
  cross.  \label{contourplot}}
\end{figure*}

\subsection{Calculation of Blazar Number and Flux}

For a given blazar, the gamma-ray flux observed on Earth is
\begin{equation} 
  F_{\gamma}(z,P)=\frac{1+z}{4\pi d_{L}(z)^{2}}\int_{E_{\rm min,obs}(1+z)/h}^{\infty}d\nu\frac{L_{\nu}(\nu,P)}{h\nu},
\end{equation}
where $d_{L}$ is the luminosity distance, $P$ is the bolometric
luminosity, and $E_{\rm min,obs}=100\rm\ MeV$ is the minimum
observable photon energy on Earth by the Fermi-LAT. 

With the GLF and SED, the number count of blazars detected above a
sensitivity $F_{\gamma}$ is
\begin{equation}\label{NgtF}
  N(>F_{\gamma})=4\pi\int_{0}^{z_{\rm max}}dz\frac{dV}{dz}\int_{L_{\gamma}^{\rm lim}(z,F_{\gamma})}^{\infty}dL_{\gamma}\rho_{\gamma}(L_{\gamma},z),
\end{equation}
where $L_{\gamma}^{\rm lim}$ is the luminosity below which a blazar at
redshift $z$ is no longer detectable for the sensitivity
$F_{\gamma}$. We set the parameter $z_{\rm max}=5$, but this does not
affect the calculation significantly, since the peak distribution is
at redshift of order unity.

The diffuse flux coming from unresolved blazars is given by
\begin{eqnarray}
  \frac{dN}{dE_{\gamma 0}dAdtd\Omega}=&&\frac{1}{4\pi}\int_{0}^{z_{\rm
      max}}dz\frac{d\chi}{dz}e^{-\tau(z,E_{\gamma 0})}\nonumber\\
  &&\times\int_{L_{\gamma,\rm min}}^{L_{\gamma}^{\rm lim}(F_{\gamma},z)}dL_{\gamma}\frac{\rho_{\gamma}(L_{\gamma},z)}{h}\nonumber\\
  &&\times\frac{L_{\nu}[E_{\gamma}/h,P(L_{\gamma})]}{E_{\gamma}}.
\end{eqnarray}
Here, $E_{\gamma}$ is the emitted photon energy [and $E_{\gamma 0} =
  E_\gamma/(1+z)$ is the observed photon energy at Earth], $A$ is area
on Earth, $t$ is time on Earth, and $\Omega$ is solid angle in the
sky. Here $L_{\nu}/(E_{\gamma})$ is the number of photons emitted per
rest-frame frequency per rest-frame time per blazar ($h$ is Planck's
constant). The quantity $dL_{\gamma}\rho_{\gamma}$ is the number of
blazars per comoving volume. The integral $d\chi$ is the line-of-sight
integral over the comoving distance. Because for $\gamma_{1}>1$ the
integral diverges at zero luminosity, $L_{\gamma,\rm min}$ is a lower
bound on the luminosity integral. We choose $L_{\gamma,\rm
  min}=10^{42}\rm\ erg\ s^{-1}$ which is an order of magnitude lower
than any Fermi-LAT observed blazar~\cite{Abdo10a,Abdo09a}. That is, we
impose a step-function cutoff of blazar GLF.  The final result is not
strongly dependent on the value of this cutoff, with a
2-order-of-magnitude difference in $L_{\gamma,\rm min}$ modifying our
best-fit parameters by $\sim$25\%.

The $\exp(-\tau)$ factor in the diffuse flux calculation accounts for
absorption of the photons on intergalactic background radiation before
reaching Earth. We use the absorption factor from Gilmore et
al.~\cite{Gilmore09}. This absorption factor was determined through
the use of galaxy formation models to find the contribution of
starlight to the absorption, as well as a contribution from quasars
which is calculated based on empirical data. This model predicts lower
values of the opacity $\tau$ than previous estimates, which leads to
less expected absorption. This is consistent with the Fermi-LAT
observing several high-energy photons coming from fairly high
redshifts, and this opacity is consistent with the findings of
Ref.~\cite{Abdo10e}.

\subsection{DGRB Spectrum Calculation}

In addition to the blazar contribution to the DGRB flux, we also
include a nonblazar AGN component to our DGRB spectrum
calculation. Ref.~\cite{Inoue08} has shown that nonblazar AGN can
account for the background radiation down to keV energies. The
combination of blazars with nonblazar AGN gives a unified model that
can explain the diffuse high-energy x-ray to gamma-ray background over
8 orders of magnitude in energy.

The AGN model we use is the model of Ref.~\cite{Inoue08}. This model
assumes the usual thermal electrons from AGN coronae, but it includes
a high-energy nonthermal component as well. These electrons
Comptonize, which produces the known x-ray spectra of AGN. This
high-energy component is analogous to the emission from solar coronae
in solar flares. Such electrons are assumed to have a power-law
injection spectrum $dN/dE\propto E^{-\Gamma}$. By adding this
nonthermal electron source to the usual thermal one, it is found that
the model matches the diffuse background spectrum well from energies
from keV to tens of MeV. 

\begin{figure*}[t]
\begin{center}
\includegraphics[width=6.8truein]{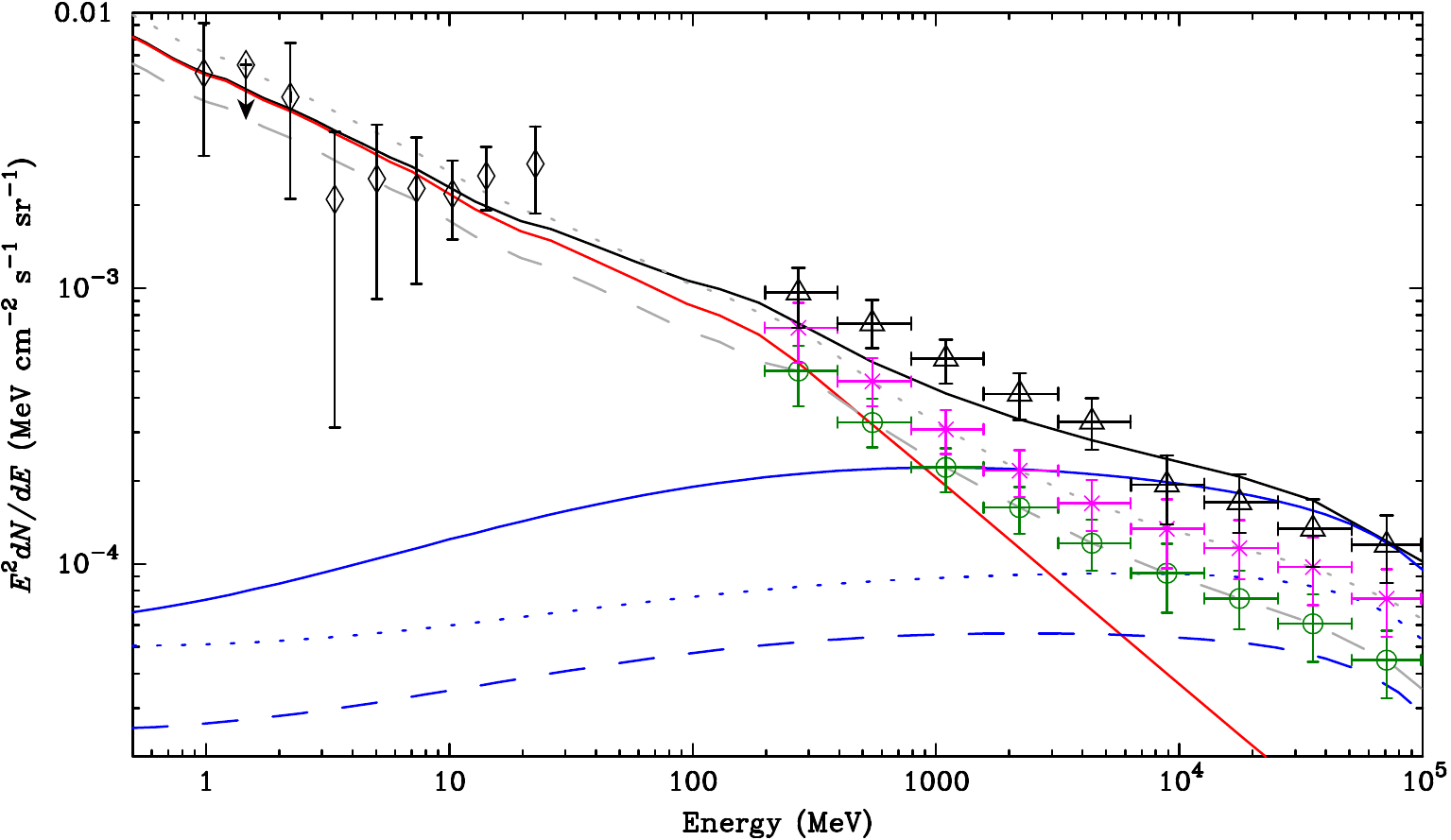}
\end{center}
\caption{Shown are the best-fit model for the current DGRB spectrum
  (solid black line) and our upper/lower $95\%$ CL forecast for the
  Fermi-LAT 5-year sensitivity (magenta-star/green-circle points). The
  low-energy-dominating solid red line is the AGN flux from
  Ref.~\cite{Inoue08}. The high-energy-dominating blue lines are the
  blazar contribution to the DGRB for the current (solid), and
  predictions for the most-optimistic (dashed) and least-optimistic
  (dotted) $95\%$ CL 5-year Fermi-LAT resolved fractions. The grey
  lines are the combined $95\%$ CL AGN plus blazar predicted flux for
  the corresponding blazar contribution. The DGRB data (triangles) are
  from FS10 and the COMPTEL data (diamonds) are from
  Ref.~\cite{Watanabe03}.\label{DGRBfitandpredict}}
\end{figure*}

Specifically, we choose the $\Gamma=3.5$ nonblazar AGN model of
Ref.~\cite{Inoue08}, which we increase in amplitude by a factor of 2
in order to match 50\% of the amplitude of the lowest-energy point in
the Fermi-LAT DGRB spectrum, with a broken power law matching the
measurements of the diffuse background by the COMPTEL Collaboration
\cite{Kappadath96}.  The power-law slope of the nonblazar AGN
spectrum is fixed by modeling of the hard x-ray luminosity function
from x-ray AGN surveys~\cite{Inoue08,Ueda:2003yx}, and the amplitude
is fixed to match the lowest point in the Fermi-LAT DGRB spectrum.
This amplitude is fixed throughout our fitting.  In order to reflect
the uncertainty of the amplitude of the flux in the lowest-energy bin,
we allow for it to have an amplitude uncertainty of 10\%, which we
vary and show in Fig. \ref{DGRBfitandpredict}.  Another low-energy
emission source, such as millisecond pulsars or star-forming galaxies,
may be responsible for the lowest-energy portion of the DGRB, but our
analysis is not strongly dependent on the spectral shape taken by the
low-energy emission source. For example, the gamma-ray spectrum from
star-forming galaxies in Ref.~\cite{Fields10} has a similar shape and
potential amplitude as the nonblazar AGN component.

In our blazar model, there are three free parameters, in addition to
those fixed in the nonblazar AGN model, as described in
Sec.~\ref{GLFsection}: $q$, $\gamma_1$, and $\kappa$. All other
parameters in the blazar model are fixed to values based on data
from other observations such as the SED sequence.  It is the purpose
of this paper to determine how well unresolved blazars can reproduce
the DGRB. Therefore, we simultaneously fit to the blazar source-count
distribution $dN/dF$ from Ref.~\cite{Abdo10} and the DGRB spectrum
from FS10. This simultaneous fit allows some freedom in the blazar
spectrum while still conforming to known blazar number
distributions. We can use the results of such a fit to constrain
models of the DGRB from unresolved blazars and predict a consistent
model of the 5-year Fermi-LAT measurements of the DGRB.

\begin{figure}[t]
\begin{center}
\includegraphics[width=3.4truein]{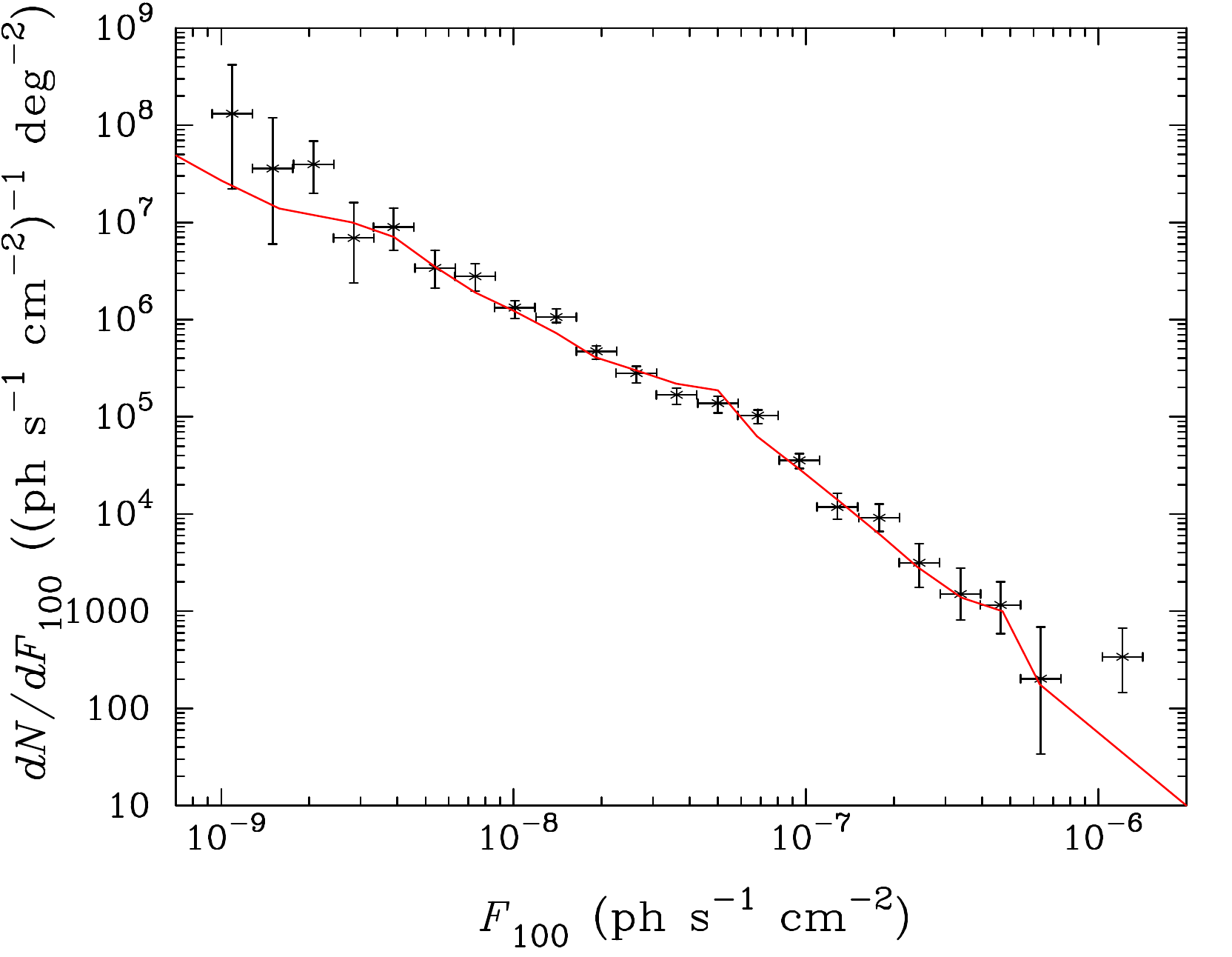}
\end{center}
\caption{Shown is the best-fit model for the source-count distribution function
  $dN/dF$ (solid line). The data are from
  Ref.~\cite{Abdo10}\label{fitteddNdF}}
\end{figure}

Fitting the model to the blazar $dN/dF$ and the DGRB spectrum, we
found that a simultaneous fit was quite reasonable. We set the lowest
blazar luminosity as $L_{\gamma,\rm min}=10^{42}\rm\ erg\ s^{-1}$, as
discussed above. The best-fit values we get are
$q=4.19^{+0.57}_{-0.13}$, $\gamma_{1}=1.51^{+0.10}_{-0.09}$, and
$\log_{10}(\kappa/10^{-6})=0.38^{+0.15}_{-0.70}$ (95\% CL). The best-fit
68\% and 95\% CL regions for $q$ and $\gamma_{1}$
are shown in Fig.~\ref{contourplot}.  These are consistent with
previous work \cite{Inoue09}, though more constrained because we are
also fitting the source-count distribution function $dN/dF$.  The
model reproduces the DGRB and blazar $dN/dF$, with a reduced
$\chi^2/{\rm DOF} = 0.63$.  The value of $q$ indicates that the
bolometric luminosity of a blazar jet is roughly 15 thousand times
more luminous than the x-ray from the accretion disk. Here,
$\gamma_{1}>1.0$ so low-luminosity blazars have significant
contributions to the total blazar flux. Therefore, a ten or more
order-of-magnitude lower value of $L_{\gamma,\rm min}$ would modify
the calculation considerably, though no blazars have been detected
below our $L_{\gamma,\rm min}$ threshold, and therefore it seems
unlikely that there is a large population of very-low-luminosity
blazars. The fraction $\kappa\simeq 2.4\times10^{-6}$ implies that
there is roughly one blazar for every 420 thousand nonblazar AGN. Our
fit to the DGRB spectrum is shown in Fig.~\ref{DGRBfitandpredict} and
the fit to $dN/dF$ is in Fig.~\ref{fitteddNdF}.

Our value for the AGN XLF and blazar GLF ratio $\kappa$, $3.4 \times
10^{-6}$ to $5 \times 10^{-7}$ (at 95\% CL), is similar to and
slightly larger than the central value derived by Inoue \& Totani
\cite{Inoue09}, $1.7 \times 10^{-6}$.  This implies that only a small
fraction of x-ray loud AGN is visible as gamma-ray blazars.  The
intrinsic jet opening angle of a blazar has been found to be $\sim$ 1
deg (subtending an area of $\sim\! 2\times 10^{-4}$ steradian)
\cite{Pushkarev:2009dx}.  Following from this is that only $\sim\!
2\times 10^{-5}$ of the AGN jets are potentially visible as blazars.
Our model then requires that only $\lesssim$ 20\% of AGN jets are
gamma-ray blazars.  This is not inconsistent with jet models
\cite{Krolik}, though if this fraction drops considerably (i.e.,
$\kappa$ is required to be much smaller), then it would call into
question the blazar model analyzed here.

Note that using the $dN/dF$ estimated from a power-law blazar spectrum
model is not perfect, due to the fact that the detection efficiency
estimate depends on the spectral model \cite{Abdo10}.  However,
Ref.~\cite{Abdo10} tested the $dN/dF$ dependence on the sensitivity
estimate with a non-power-law fit to the blazar spectra and found it
did not significantly change the measurement of $dN/dF$.  We also
verified this sensitivity dependence with a test fitting by increasing
the errors on the measured $dN/dF$ at low flux, and we found that our
model did not prefer a different amplitude or shape to the
source counts at the low flux where the efficiency for blazar detection
is low.

Refs.~\cite{Inoue09,Inoue10a} used a combined GLF plus SED model to predict
the Fermi-LAT's ability to observe blazars and their spectra, using
the results of the EGRET Collaboration. The paper fit its GLF
parameters using the redshift and gamma-ray luminosity distributions
of EGRET blazars. This led to a prediction that 600 to 1200 blazars
should be resolved in 5 years of Fermi-LAT data, which would yield $98\%$
to $100\%$ of the total blazar flux. However, the cumulative number of
blazars predicted by that paper is in disagreement with the
observations of the Fermi-LAT~\cite{Abdo10}. The cumulative number
count by Ref.~\cite{Inoue09} is predicted to have a break at
$~10^{-7}\rm\ photons\ cm^{-2}\ s^{-1}$ whereas the break seen by the
Fermi-LAT Collaboration is at $~5\times10^{-8}\rm\ photons\ cm^{-2}\
s^{-1}$. Also, the surface density of sources predicted in that paper
is too small to match the measured value.

Importantly, Refs.~\cite{Inoue09,Inoue10a} fit their model to the
EGRET catalog blazar spectra SED, not that from Fermi-LAT. The EGRET
telescope had strong cuts which limited high-energy photon
observations, which lead to EGRET only observing a few BL
Lacs~\cite{Abdo10}. Also, the redshift and luminosity distributions
are strongly dependent on detector sensitivity, because BL Lacs have
lower luminosity and therefore are observed at lower redshifts. This
means that the current data for the overall blazar redshift
distribution, in particular, is more strongly biased toward lower
redshifts than the complete distribution. Ref.~\cite{Inoue10} posited
that one significant source for the difference between this
calculation and the Fermi-LAT results comes from needing to correctly
account for Fermi-LAT sensitivities. By fitting to $dN/dF$, which is
not as heavily dependent on detector sensitivity, we can get a more
robust prediction that should not change significantly for different
sensitivities.  Refs.~\cite{Inoue09,Inoue10a} argued that a model of
this type should roughly match the DGRB spectrum. In
Ref.~\cite{Inoue09}, the model parameters were fit to the EGRET DGRB
spectrum, and, as discussed above, the model parameters are roughly
consistent with our results.  In our analysis here, we use the DGRB
spectrum and flux source counts, as measured by the Fermi-LAT, as a
constraint in order to determine how well this class of models fits
the DGRB and blazar population. For those models that fit the
spectrum, we can determine the predicted values for the DGRB flux at
the Fermi-LAT's 5-year sensitivities and determine the theoretical
uncertainty on these predictions.

In another analysis of the contribution of blazars to the DGRB, the
Fermi-LAT Collaboration used the currently measured differential
number distributions of blazars ($dN/dF$) and blazar gamma-ray index
($\Gamma$) distributions to estimate the contribution of unresolved
blazars to the DGRB~\cite{Abdo10}. In that analysis, it was found that
less than 20\% of the DGRB can be accounted for by blazar
emission. However, in that calculation, the assumption was made that
the distribution of indices $\Gamma$ is independent of
sensitivity. Because less-luminous BL Lacs have significantly
different indices than more luminous FSRQs, the overall distribution
of indices should change as better sensitivity allows a greater
fraction of BL Lacs to be detected.

Additionally, it was shown in Refs.~\cite{Abdo10a,Abdo09a} that a
basic power-law model does not fit the individual blazar spectra well,
especially for the low-luminosity BL Lacs. A GLF plus SED model should
overcome these issues. The GLF accounts for differing redshifts of
blazars, so the relationship between flux sensitivity and luminosity
detectability is well-defined. The SED accounts for the distribution
of luminosities with energy, so a calculation around the IC peaks for
BL Lacs should more realistically reproduce the contribution to the
DGRB from blazars than a simple distribution of photon indices. This
is especially important to incorporate when determining the
contribution of unresolved low-luminosity blazars to the DGRB, since
they have much harder spectra than high-luminosity blazars.

\section{5-year Predictions for Blazars and the DGRB}

\begin{figure}[t]
\begin{center}
\includegraphics[width=3.4truein]{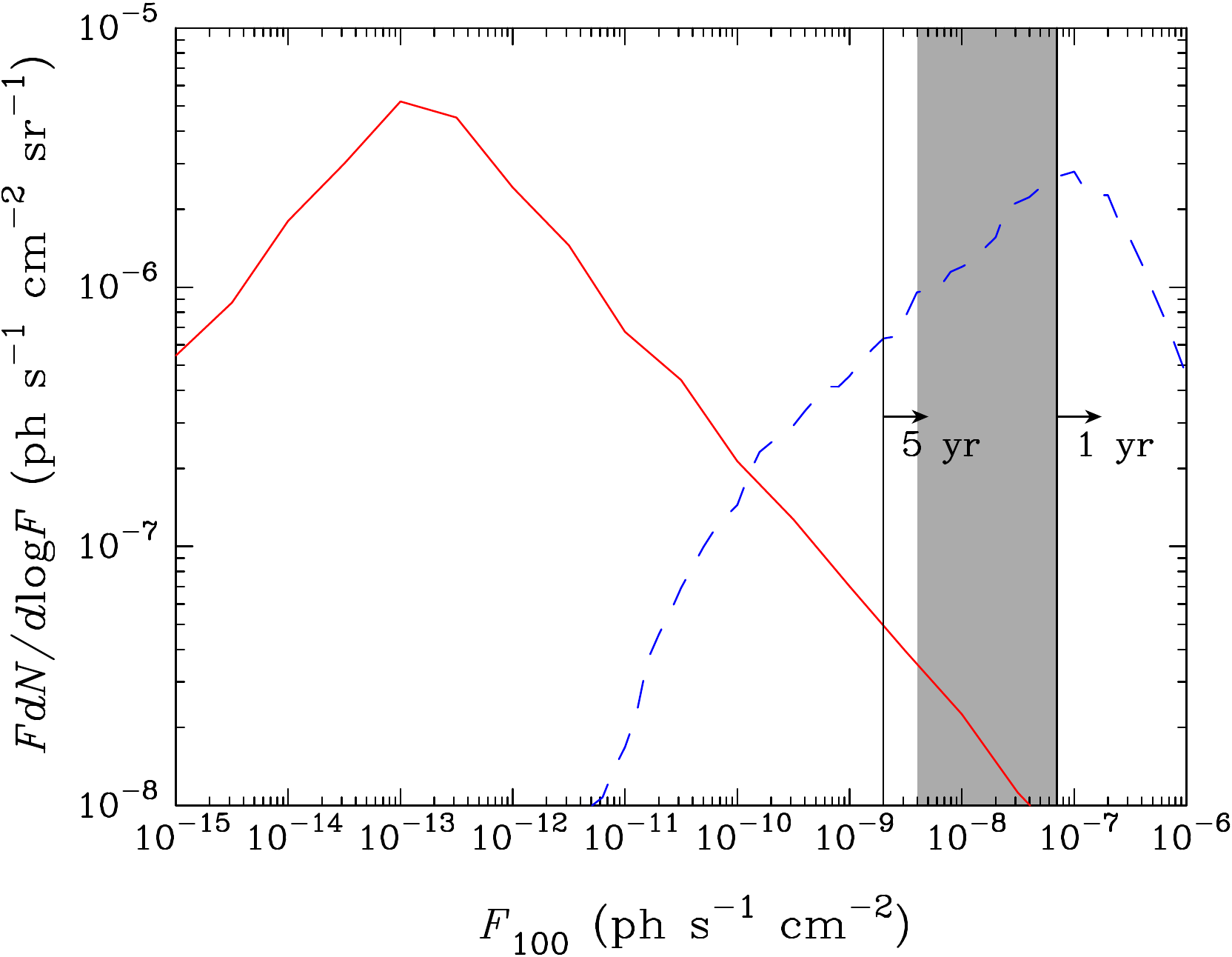}
\end{center}
\caption{Shown is the flux per logarithmic sensitivity for our
  best-fit model. The dashed line is the flux coming from blazars and
  the solid line is the flux coming from nonblazar AGN. The vertical
  solid lines with arrows mark the sensitivity to {\it all}
  spectral-index sources at the Fermi-LAT 1-year and the projected
  5-year sensitivity of Fermi-LAT.  The gray boxed region indicates
  the range of sensitivity at 1-year to sources with the spectral
  indices of the bulk of the blazar population, as in
  Fig.~\ref{point}. \label{FdNdlogF}}
\end{figure}

We adopt the 5-year predictions for a sensitivity to point-sources by
Fermi-LAT of $S_5=2\times 10^{-9}\rm\ photons\ cm^{-2}\ s^{-1}$ above
$100\rm\ MeV$. This value is consistent with the Fermi-LAT
Collaboration's estimate of the LAT sensitivity to point sources with
gamma-ray index of $\sim$2 \cite{Atwood09}.\footnote{\tt
  http://fermi.gsfc.nasa.gov/science/433-SRD-0001\_CH-04.pdf} As
discussed earlier, the majority of low-flux blazars are expected to be
BL-Lacs, which predominantly have radio luminosity less than
$10^{43}\rm\ erg/s$ \cite{Abdo10d}. Such low-luminosity blazars have gamma-ray
indices of $\sim$2 or less, according to the blazar SED. Therefore, we
find the use of $S_5=2\times 10^{-9}\rm\ photons\ cm^{-2}\ s^{-1}$ as
the Fermi-LAT 5-year sensitivity to blazars of all gamma-ray indices
to be a reasonable estimate.

To determine the total number of blazars detectable by the Fermi-LAT,
we need to take Eq.~(\ref{NgtF}) down to a sensitivity of
$S_{5}$. Similarly, we can determine the total number of blazars in
the sky by letting the sensitivity go to zero flux. With $95\%$ CL, we
predict that there are $5.4^{+1.8}_{-1.7}\times 10^4$  total blazars in the
observable universe. Of these, $2415^{+240}_{-420}$  should be
detectable by the Fermi-LAT after 5 years of running. The amount of
flux coming from blazars per logarithmic sensitivity is shown in
Fig.~\ref{FdNdlogF}. Our prediction is that $94.7^{+1.9}_{-2.1}\%$ 
of blazar flux is expected to be resolved by the Fermi-LAT after 5
years, mostly at lower energies. In contrast, the flux for nonblazar
AGN should not be appreciably resolved for another 4 orders of
magnitude in sensitivity.

\begin{figure}[t]
\begin{center}
\includegraphics[width=3.4truein]{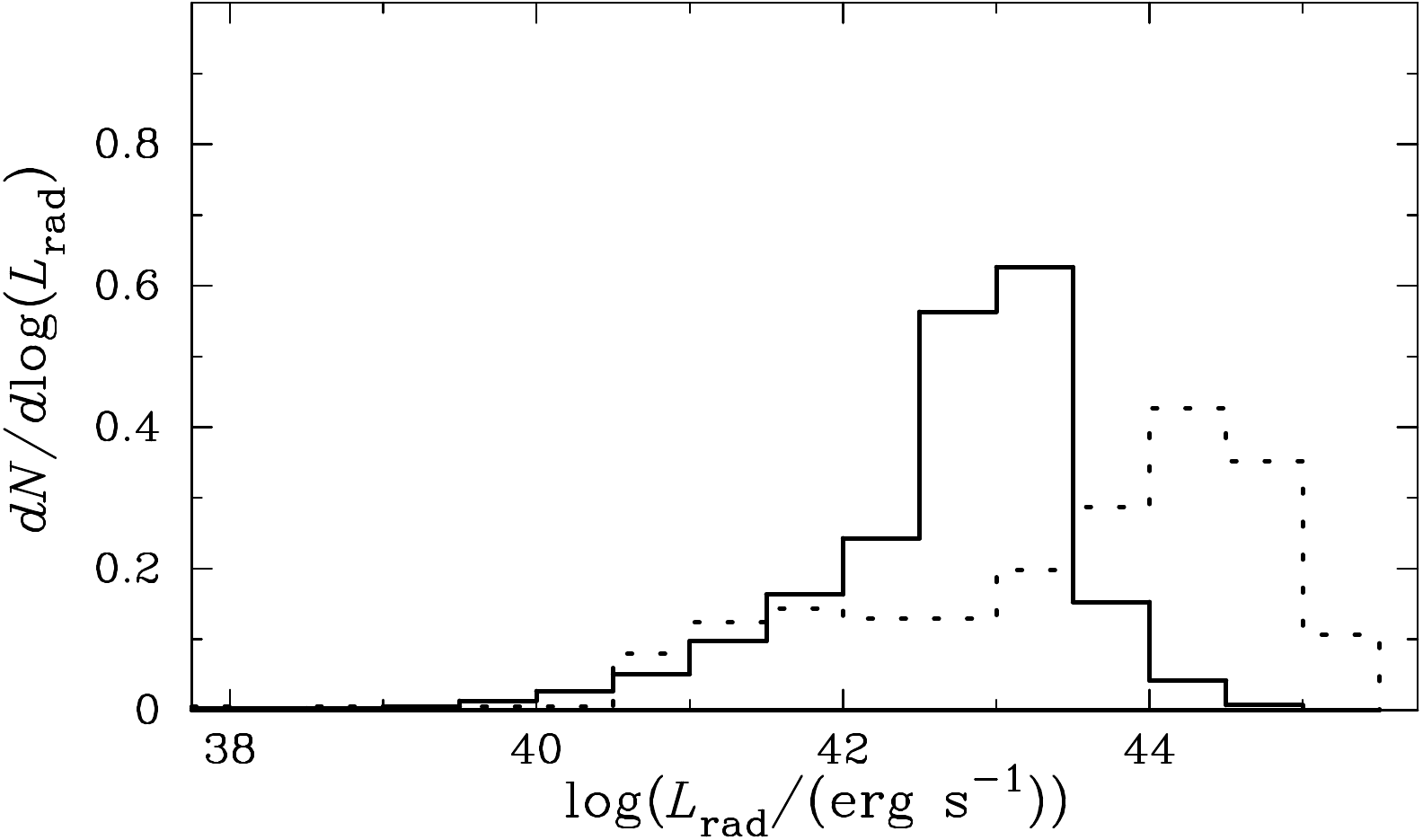}
\end{center}
\caption{Shown is the radio luminosity distribution of blazars. The
  solid line is our prediction for the distribution after 5 years of
  Fermi-LAT running. The dotted line is the current Fermi-LAT
  distribution for blazars~\cite{Abdo10d}. Each distribution is
  independently normalized to unity.\label{lraddistprediction}}
\end{figure}

\begin{figure}[t]
\begin{center}
\includegraphics[width=3.4truein]{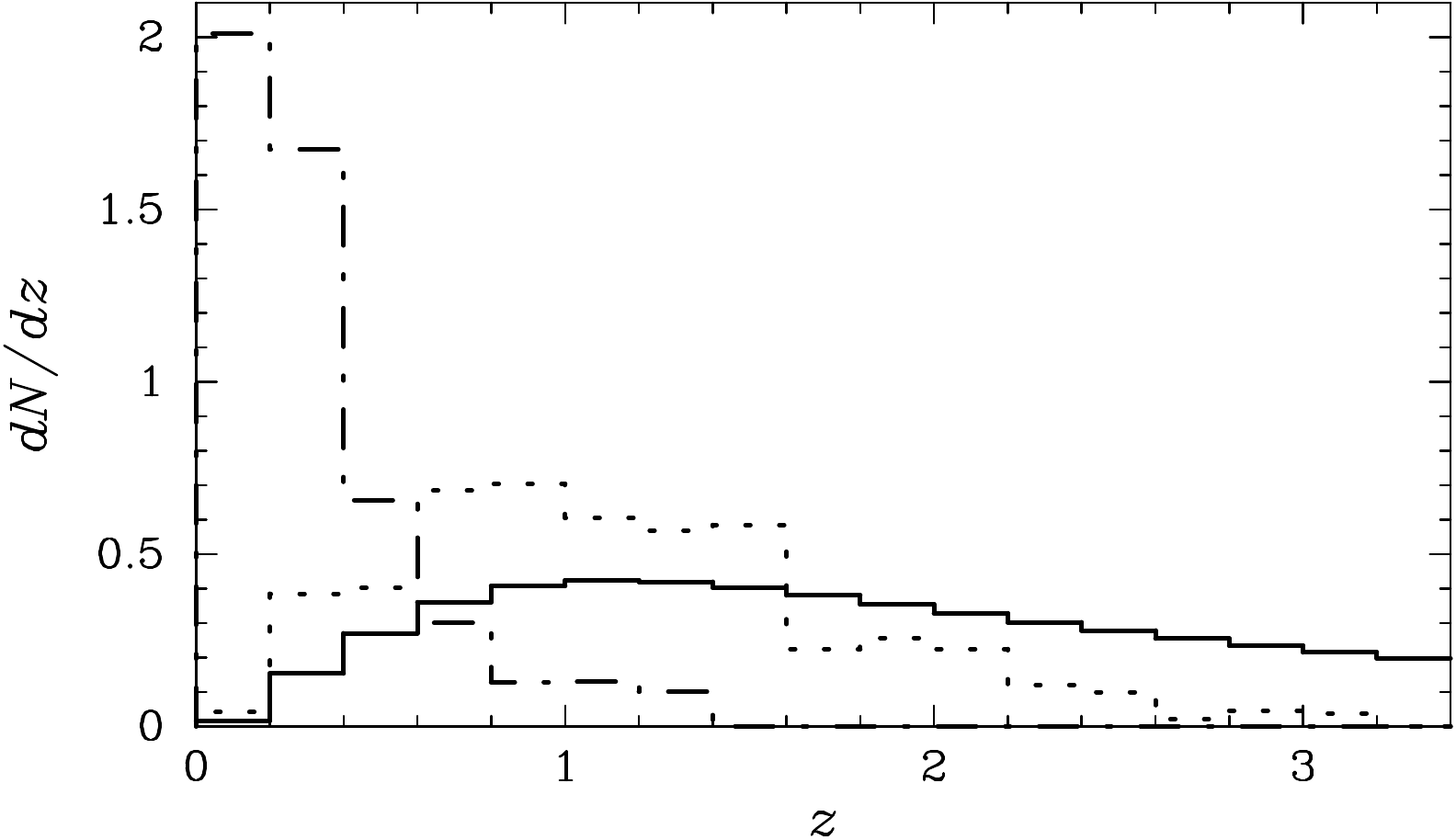}
\end{center}
\caption{Shown are the distribution in redshift of blazars. The solid
  line is our prediction for the distribution after 5 years of LAT
  running. The dotted line is the current Fermi-LAT-measured
  distribution for FSRQs and the dot-dashed line is the current
  Fermi-LAT-measured distribution for BL Lacs~\cite{Abdo10d}. Each
  distribution is independently normalized to
  unity.\label{zdistprediction}}
\end{figure}

In addition to the number counts of blazars, we can also predict the
distributions of blazars in luminosity and redshift. To get these
distributions, we differentiate Eq.~(\ref{NgtF}). The distribution of
blazars in radio luminosity, shown in Fig.~\ref{lraddistprediction},
shifts toward lower luminosities at better sensitivities. This is due
to the FSRQ population being mostly resolved, whereas the new resolved
sources at better sensitivities are mostly low-luminosity BL Lacs. The
redshift distribution of blazars, Fig.~\ref{zdistprediction}, should
shift toward higher redshifts as sensitivity improves. Because the
FSRQ sample is mostly complete, it would be expected that the redshift
distribution of BL Lacs, and blazars in general, should be roughly
similar to the current redshift distribution of FSRQs. Our prediction
of the redshift distribution of blazars after 5 years of Fermi-LAT
running matches well with the current FSRQ distribution, which
provides a verification of our theory and fit parameters.  Note that
the FSRQ sample is not totally complete, and the objects to be
resolved at $z\gtrsim 2$ would be FSRQs.  As can be seen in Fig. 9
of Ref.~\cite{Abdo10d}, the distribution of FSRQs reaches the current
flux limit, so there remains a population of high-luminosity, soft
spectral index, high redsift FSRQs to be resolved.

In our model, we fit the total blazar plus AGN flux to the DGRB
spectrum for the spectrally-dependent sensitivity as described
above. The model fit worked exceptionally well, indicating that a
combination of blazar flux with the flux of nonblazar AGN makes up
all the DGRB over a wide range in energies. With this fit, we then
calculated what the combined flux should be after 5 years of Fermi-LAT
observations, giving the sensitivity of $2\times 10^{-9}\rm\ photons\
cm^{-2}\ s^{-1}$. The upper and lower bounds of the $95\%$ CL region
of this calculation are given by the upper and lower forecast points
in Fig.~\ref{DGRBfitandpredict}. We have included a $10\%$ uncertainty
on the nonblazar AGN flux in this error estimate to account for the
error in the lowest-energy bins' constraint on the AGN model. At 100
GeV, we expect the DGRB to decrease by a factor of 1.6 to 2.6 at the
95\% CL upper and lower flux limits, whereas at 100 MeV the DGRB only
decreases by a factor of 1.3 to 1.9. The difference in DGRB
improvement is due to a greater fraction of the DGRB being due to
blazars at high energies, while the nonblazar AGN flux dominates at
low energies.  Importantly, the resolution of sources can do better
than the square root of exposure time due to the increased prevalence
of easily-detected hard sources beyond, but near, the current
point-source flux-limit sensitivity.

\section{Conclusions}

We have shown that the DGRB can be composed entirely by gamma-rays
produced in blazars and nonblazar AGN.  The LDDE plus SED-sequence is
a physical model for the spectral evolution of a
cosmologically-evolving blazar population contributing to the DGRB
based on the unified AGN model for blazars.  This model successfully
accounts for the full DGRB spectrum as well as the full blazar
source-count distribution function, which, unlike other approaches,
are not used as components of the model.  Independent of the nonblazar
AGN component, the blazar model produces nearly the entire DGRB at its
highest measured energies.  The small value of $\kappa\simeq
2.4\times10^{-6}$, the x-ray AGN fraction seen as blazars, constrains
this model to require a small fraction, $\lesssim$20\%, to be both
properly oriented and sufficiently energetic in order to be gamma-ray
emitters.

We found constraints on this model from the spectrum of the DGRB and
source-count distribution function $dN/dF$ of blazars as observed by
Fermi-LAT.  Our results are consistent with previous work by Inoue \&
Totani \cite{Inoue09} which employed EGRET spectral data to forecast
the Fermi-LAT DGRB.  We forecast that $94.7^{+1.9}_{-2.1}\%$ of the
flux from blazars will be resolved into point sources by Fermi-LAT
with 5 years of observation, with a corresponding reduction of the
flux in the DGRB by a factor of $\sim$2 to 3 (95\% CL) from the
automatic removal of these sources in the measurement of the DGRB.  
This has significant consequences for the sensitivity of the DGRB
measurement to dark matter annihilation, which we explore in a
companion paper~\cite{Abazajian10a}.

We predict that $2415^{+240}_{-420}$ blazars should be resolved, of
$5.4^{+1.8}_{-1.7}\times 10^4$ total blazars in the universe (95\%
CL).  Recent results of anisotropy in the DGRB also indicate the
likely presence of an unresolved point-source
population~\cite{Vargas:2010en}.  Using tests with enhanced
point-source sensitivity, we find that future gamma-ray experiments at
Fermi-LAT energies will resolve the blazar contribution to the DGRB
such that the flux in the DGRB decreases as the square root of the
point-source sensitivity.

The LDDE plus SED-sequence model is more complex than the
over-simplistic source-count method with a fixed spectral-index
distribution adopted by the Fermi-LAT Collaboration in FB10, yet it
has {\it fewer} free parameters for the blazar population than the
more simplified model (three versus four free for the blazar model,
plus those fixed in the nonblazar AGN model in this work).  Most
importantly, the Fermi-LAT analysis of FB10 fixes the spectral index
of the blazar population, and, crucially, does not include the
hardening of the spectra of the unresolved low-luminosity blazar
population.  The hardening of spectra with lower luminisity has been
seen by both EGRET~\cite{Fossati97,*Fossati98,*Donato01} and Fermi-LAT
(Fig.~\ref{point}).  The fixed spectrum forces the FB10 conclusion
that only $\sim$16\% of the GeV isotropic diffuse background could
arise from blazars, and is also the case in other work using fixed
blazar spectra \cite{Malyshev:2011zi}.  Other recent work with
different blazar population models, including spectral shape
variation~\cite{Venters:2011gg}, possible point-source
confusion~\cite{Stecker:2010di}, and BL Lac dominance of the
unresolved portion~\cite{Neronov:2011kg} also find that a substantial
portion of the DGRB could arise from the blazar population.

Overall, the SED-sequence model of blazars and AGN as the source of
the DGRB is remarkably consistent with the measured DGRB spectrum and
blazar source-count distribution.  The SED-sequence will continue to
be improved with upcoming Fermi-LAT blazar data~\cite{Meyer:2011uk}.  Further
analyses of the type presented here, incorporating potential
enhancements to the SED-sequence model, the XLF of AGN, and general
studies of observed blazar spectral properties, will further enlighten
the understanding of the extragalactic gamma-ray sky.

\begin{acknowledgments}
  We would like to thank P.\ Agrawal, M.\ Ajello, J.\ Beacom,
  Z.\ Chacko, D.\ Malyshev and J.\ McEnery for useful discussions. KNA
  and JPH are supported by NSF Grant No.\ 07-57966 and NSF CAREER
  Grant No.\ 09-55415. SB has been partially supported by MICNN,
  Spain, under FPA Contract No.\ 2007-60252 and Consolider-Ingenio
  CPAN CSD2007-00042 and by the Comunidad de Madrid through HEPHACOS
  Project No.\ ESP-1473. SB acknowledges support from the CSIC Grant
  No.\ JAE-DOC.
\end{acknowledgments}

\appendix

\section{Blazar SED Sequence}\label{SED Appendix}

The full SED fit is given as a function of radio luminosity $\psi_{R}$
and the logarithm of rest-frame frequency $x$. We follow
Ref.~\cite{Inoue09} in the formulation of the SED. The radio
luminosity is used to distinguish between SEDs for blazars of
different bolometric luminosity. This separation of SED by total
luminosity should account for the difference in spectral index seen by
the Fermi-LAT between higher-luminosity FSRQs and
lower-luminosity BL Lacs~\cite{Ghisellini09,Abdo09a,Abdo10a}.
\begin{eqnarray}
x&\equiv&\log_{10}(\nu/\rm Hz),\\
\psi(x;\psi_{R})&\equiv&\log_{10}[\frac{\nu L_{\nu}(\nu(x),P(\psi_{R}))}{\rm erg\ s^{-1}}],\\
\psi_{R}&\equiv&\psi(x=9.698) .
\end{eqnarray}
The full model is the sum of a synchrotron [$\psi_s(x)$] and inverse
Compton [$\psi_c(x)$] component.
\begin{equation}
\psi(x)=\log_{10}[10^{\psi_s(x)}+10^{\psi_c(x)}] .
\end{equation}

Each component is parameterized as the sum of a lower-frequency linear
part and a higher-frequency parabolic part. Here, $x_{tr,s}$ and
$x_{tr,c}$ are the frequencies where the linear part transitions to
the parabolic part for the synchrotron and IC component. The linear
parts are written as 
\begin{eqnarray}
\psi_{s1}(x)&\equiv&(1-\alpha_{s})(x-x_{R})+\psi_{R}\ (x<x_{tr,s}),\\
\psi_{c1}(x)&\equiv&(1-\alpha_{c})(x-x_{X})+\psi_{X}\ (x<x_{tr,c}),
\end{eqnarray}
where $\alpha_{s}=0.2$ and $\alpha_{c}=0.6$ are the
$L_{\nu}\propto\nu^{\alpha}$ indices in the radio and hard x-ray
bands, respectively. The characteristic radio and hard x-ray
frequencies are $x_{R}=9.698$ and $x_{X}=17.383$. The radio luminosity
$\psi_{R}$ is an input parameter to the theory and the hard x-ray
luminosity is fitted to the data as
\begin{equation}
\psi_{X}=
  \left\{ 
    \begin{array}{rl} 
      (\psi_{R}-43)+43.17 &\ \psi_{R}\leq 43 \\
      1.40(\psi_{R}-43)+43.17 &\ 43<\psi_{R}\leq 46.68 \\
      1.40(46.68-43)+43.17 &\ \psi_{R}>46.68\ .
    \end{array} 
    \right.
\end{equation}
The parameter $\psi_{X}$ is kept constant for $\psi_{R}>46.68$ because
the continuity of the IC component cannot be satisfied above this
value. However, this hard x-ray luminosity corresponds to a gamma-ray
luminosity well above the maximum detected gamma-ray luminosity, so it
does not affect the calculation of the DGRB.

The parabolic parts of the components are parameterized as
\begin{eqnarray}
\psi_{s2}(x)&\equiv&\psi_{s,p}-[(x-x_{s})/\sigma]^{2}\ (x\geq x_{tr,s}),\\
\psi_{c2}(x)&\equiv&\psi_{c,p}-[(x-x_{c})/\sigma]^{2}\ (x\geq x_{tr,c}),
\end{eqnarray}
where $x_{s}$ and $x_{c}$ are the synchrotron and IC peak frequencies,
$\psi_{s,p}$ and $\psi_{c,p}$ are the synchrotron and IC peak
luminosities, and $\sigma$ is the width of the parabolas.

By requiring continuity of the synchrotron component from the
linear-to-parabolic parts, we have
\begin{equation}
\psi_{s,p}=(1-\alpha_{s})(x_{tr,s}-x_{R})+\psi_{R}+\left(\frac{x_{tr,s}-x_{s}}{\sigma}\right)^{2}\ .
\end{equation}
Similarly, the continuity of the IC component gives
\begin{eqnarray}
x_{tr,c}&=&\frac{-\zeta-\sqrt{\zeta^{2}-4\eta}}{2},\\
\zeta&=&\sigma^{2}(1-\alpha_{c})-2x_{c},\\
\eta&=&x_{c}^{2}+\sigma^{2}[\psi_{X}-x_{X}(1-\alpha_{c})-\psi_{c,p}]\ .
\end{eqnarray}
By inspection
\begin{eqnarray}
&&x_{tr,s}=10.699,\\
&&x_{c}=x_{s}+8.699\ .
\end{eqnarray}
Fitting to data, the rest of the parameters are given by 
\begin{eqnarray}
x_{s}&=&
  \left\{ 
    \begin{array}{ll} 
       -0.88(\psi_{R}-43)+14.47 &\ \ \ \psi_{R}\leq 43 \\
       -0.40(\psi_{R}-43)+14.47 &\ \ \ \psi_{R}>43 \\
    \end{array} 
    \right. \\
\sigma&=&
  \left\{ 
    \begin{array}{ll} 
      0.0891 x_{s}+1.78 &\ \psi_{R}\leq 43 \\
      \left[ 2(x_{s}-x_{tr,s})/(1-\alpha_{s})\right]^{1/2} &\ \psi_{R}>43 \\
    \end{array} 
    \right. \\
\psi_{c,p}&=&
  \left\{ 
    \begin{array}{ll} 
      \psi_{s,p} &\ \ \psi_{R}\leq 43 \\
      1.77(\psi_{R}-43)^{0.718}+45.3 &\ \ \psi_{R}>43\ .\\
    \end{array} 
    \right.
\end{eqnarray}
These parameters have been chosen such that the luminosity changes
continuously with $\psi_{R}$ over all luminosities and to make the
synchrotron linear-to-parabolic transition smooth for large
$\psi_{R}$.

\section{X-ray Luminosity Function}\label{XLF Appendix}

The x-ray luminosity function $\rho_{X}$ is the comoving number
density of AGN per unit x-ray AGN disk luminosity $L_{X}$. The model
of Refs.~\cite{Ueda03,Hasinger05} give the distribution as
\begin{equation}
\rho_{X}(L_{X},z)=\rho_{X}(L_{X},0)f(L_{X},z).
\end{equation}
The present distribution is given by
\begin{equation}
  \rho_{X}(L_{X},0)=\frac{A_{X}}{L_{X}
    ln(10)}\left[\left(\frac{L_{X}}{L_{X}^{*}}\right)^{\gamma_{1}} + 
    \left(\frac{L_{X}}{L_{X}^{*}}\right)^{\gamma_{2}}\right]^{-1}.
\end{equation}
The density evolution is given by
\begin{equation}
f(L_{X},z)=
  \left\{ 
    \begin{array}{ll} 
      (1+z)^{p_{1}}  & z\leq z_{c}(L_{X}) \\
      (1+z_{c}(L_{X}))^{p_{1}}\left(\frac{1+z}{1+z_{c}(L_{X})}\right)^{p_{2}} &  z > z_{c}(L_{X}).
    \end{array} 
    \right.  
\end{equation}
The peak evolution happens at $z_{c}$, given by
\begin{equation}
z_{c}(L_{X})=
  \left\{ 
    \begin{array}{ll} 
      z_{c}^{*}  & L_{X}\geq L_{a} \\
      z_{c}^{*}(L_{X}/L_{a})^{\alpha}  & L_{X} < L_{a}.
    \end{array} 
    \right. 
\end{equation}
The evolution indices $p_{1}$ and $p_{2}$ are
\begin{eqnarray}
  p_{1}&=&p_{1}^{*}+\beta_{1}[\log_{10}(L_{X})-44.0]\\
  p_{2}&=&p_{2}^{*}+\beta_{2}[\log_{10}(L_{X})-44.0] .
\end{eqnarray}
The parameters for the models are given in Table I.  If
$\gamma_{1}>1$, then the integrated background flux diverges, so we
set the minimum gamma-ray luminosity to $L_{\gamma,\rm
  min}=10^{42}\rm\ erg/s$. This is an order-of-magnitude lower than
any Fermi-LAT observed blazar, and the results are not sensitive to
this value being lowered slightly~\cite{Abdo10a,Abdo09a}.

\begin{center}
\begin{tabular}[t]{|l|l|l|}
  \multicolumn{3}{c}{Table 1} \\
  \hline
  \multicolumn{3}{|c|}{Parameters for the AGN XLF} \\
  \hline
  Parameter & Ueda et al. 2003 & Hasinger et al. 2005\\
  \hline
  $A_{X}\rm\ (Mpc^{-3})$ & $5.04\times10^{-6}$ & $2.62\times10^{-7}$\\
  $log_{10}L_{X}^{*}$ & $43.94_{-0.26}^{+0.21}$ & $43.94\pm0.11$\\
  $\gamma_{2}$ & $2.23\pm0.13$ & $2.57\pm0.16$\\
  $z_{c}^{*}$ & 1.9, fixed & $1.96\pm0.15$\\
  $log_{10}L_{a}$ & 44.6, fixed & 44.67, fixed\\
  $\alpha$ & $0.335\pm0.07$ & $0.21\pm0.04$\\
  $p_{1}^{*}$ & $4.23\pm0.39$ & $4.7\pm0.3$\\
  $p_{2}^{*}$ & -1.5, fixed & $-1.5\pm0.7$\\
  $\beta_{1}$ & 0.0, fixed & $0.7\pm0.3$\\
  $\beta_{2}$ & 0.0, fixed & $0.6\pm0.8$\\
  \hline
  \multicolumn{3}{|c|}{Note: Luminosities are in erg/s}\\
  \hline
\end{tabular}
\end{center}
\ \\ \ 

\bibliography{bibliography}
\end{document}